\documentclass[useAMS,usenatbib]{mn2e}
\usepackage{graphicx}
\title[Multicolour polarimetry of $\gamma$-ray binary HESS J0632+057]{First multicolour polarimetry of TeV $\gamma$-ray binary
HESS J0632+057 close to periastron passage}
\author[Yudin, Potter \& Townsend]{R. V. Yudin$^{1}$\thanks{E-mail: ruslan.yudin@gao.spb.ru}, S. B. Potter$^{2}$ and L. J. Townsend$^{3}$\\
$^{1}$ Pulkovo Observatory, Pulkovskoye sh. 65, Saint-Petersburg 196140, Russia\\
$^{2}$ South African Astronomical Observatory, PO Box 9, 7935
Observatory, Cape Town, South Africa\\
 $^{3}$ Astrophysics, Cosmology and Gravity Centre, Department of Astronomy, University of Cape Town, Rondebosch 7701, South Africa}
\begin{document}

\date{Accepted 2016 xxx xx. Received 2016 March xx; in original form 2016 March xx}

\pagerange{\pageref{firstpage}--\pageref{lastpage}} \pubyear{2016}

\maketitle

\label{firstpage}

\begin{abstract}
We present the results of $U\!BVRI$ polarimetry of the TeV
$\gamma$-ray binary HESS J0632+057 obtained on 2015 March 24 (JD
2457106) and 2015 December 12 (JD 2457369). The
detected polarisation values of HESS J0632+057, just after
periastron passage (March 24), are higher than all previously
published values ($p_{V}\sim$4.2 per cent) and the position angle
($\Theta_{obs}\sim171^{o}-172^{o}$) is also different by
$\sim6^{o}-10^{o}$ from previously published values. The data
obtained just before the subsequent periastron passage (December
12) show statistically lower polarisation in all photometric bands
($p_{V}\sim$3.9 per cent) and a different position angle
$\Theta_{obs}\sim167^{o}-168^{o}$. From observations of a nearby
field star, the interstellar component of the measured
polarisation was estimated as $p_{is}^{V}\sim$0.65 per cent and
$\Theta_{is}\sim 153^o$. This estimate was used with the previous
"V"-band estimation by "field-stars method" ($p_{is}^{V}\sim$2 per
cent and $\Theta_{is}\sim 165^o$) of \citet{b33} to identify the
wavelength dependence of the intrinsic polarisation in HESS
J0632+057. It was found that after subtraction of the interstellar
component (for both $p_{is}$ estimates), the wavelength dependence
of the intrinsic polarisation in HESS J0632+057 is essentially
flat. We propose that the formation of an additional source of
polarisation or some perturbation of circumstellar material at
this orbital phase can explain the changes in the level of
polarisation in HESS J0632+057 close to periastron passage.
\end{abstract}


\section{Introduction}
HESS J0632+057 (MWC\,148; HD259440) is one of the most interesting
TeV $\gamma$-ray binaries \citep{b8,b9,b10}. The optical star is
known to be a B0 emission line star, though the nature of the
compact companion (a neutron star or a black hole) remains
uncertain \citep{b35,b36}. The object displays periodic variations
on several different time-scales. The most pronounced variations,
derived from X-ray data, were detected with a period of
315$^{+6}_{-4}$ (or 321$\pm$5) days \citep{b20} which is
considered to be the orbital period of the binary system. Other
variability includes:
\begin{itemize}
\item H$_{\alpha}$ line variations with a period of $\sim$60$^{d}$
detected by \citet{b7}, possibly attributed to a spiral density
wave in the circumstellar disc, \item $\gamma$-ray variability on
time-scales of the order of one to two months, possibly linked to
the X-ray outburst that occurs about 100 days after periastron
passage \citep{b25}, \item clear evidence of strong X-ray
variability on multiple timescales, with the measured flux
doubling on time-scales as short as $\sim$5 days \citep{b27}.
\end{itemize}

\begin{figure*}
\includegraphics[width=170mm,angle=0]{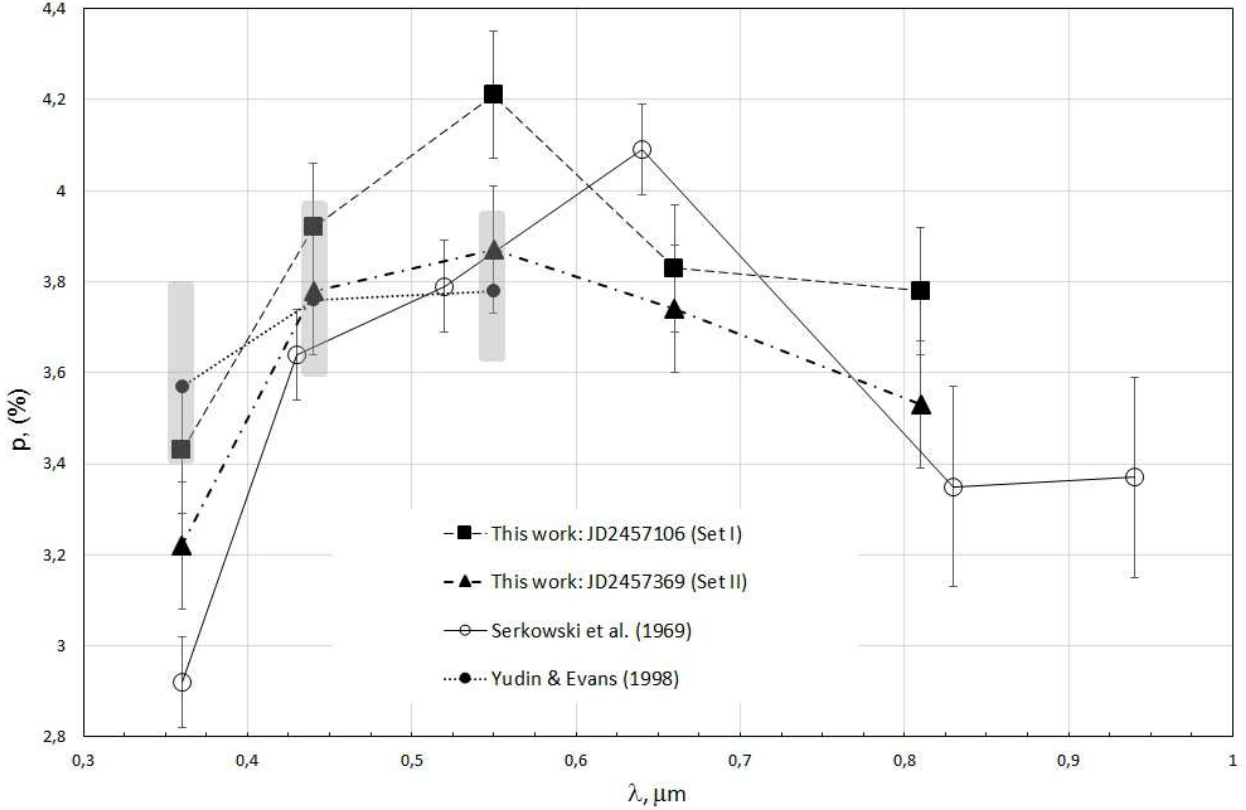}
\vspace{-0.5cm}
 \caption{Observed wavelength dependence of linear polarisation for
 HESS J0632+057 from \citet{b16} - open circles and solid line, from
 \citet{b5} - filled circles and dotted line, and from this work: JD2457106 (Set I) - filled squares and dashed line,
JD2457369 (Set II) - filled triangles and dash-dotted line. Gray
zones on the plot indicate the spread of the polarisation degree
for \citet{b5} data in corresponding filters.}
 \label{figure_1}
\end{figure*}
\begin{table*}
 \centering
  \caption{Multicolour polarimetric data of \citet{b16}. No exact JD date and $\sigma_{\Theta}$ are available. Observations were made in 1966.}
  \begin{tabular}{lccccccc}
  \hline
   $\lambda$, ($\mu$m)    & 0.33   &    0.36        & 0.43  & 0.52 & 0.64 & 0.83  & 0.94 \\

         &  &  nearly "U"       & nearly "B"  &  nearly "V"  & nearly "R" & nearly "I"  &  \\
 \hline
     p$\pm\sigma$       &   ---    &    2.92$\pm0.10$        & 3.64$\pm0.10$       & 3.79$\pm0.10$        & 4.09$\pm0.10$ & 3.35:  & 3.37: \\
     $\Theta_{obs}$      &  159.7      &    164.6        & 165.2       & 165.9        & 163.1 & 164.5:  & 166.0: \\
\hline
\end{tabular}
\end{table*}

Although the object has been intensively studied across a wide range
of the electromagnetic spectrum, polarimetric data of HESS
J0632+057 is very limited (see later discussion in Sect.~3).
There are no published multicolour polarimetric data covering the
optical region from $\sim$0.3$\mu$m to 0.9$\mu$m. The only data
published several decades ago by \citet{b16} were obtained with
large errors in the red ($\sigma_{p}>$0.2 per cent at wavelengths
$\sim$0.8-0.9$\mu$m, see Table~1). This does not allow one to
identify the true wavelength dependence of the observed HESS
J0632+057 polarisation. Moreover, without these data and estimates
of the interstellar component of polarisation in all optical
photometric bands, the wavelength dependence of intrinsic
polarisation can not be determined. Finally, taking into account
uncertainties in the orbital period determination ($\pm$5 days),
it is not possible to associate previous sparse polarimetric data
with a specific orbital phase by extrapolating the modern ephemerides
back several decades. However, we expect that the long-term
polarimetric variability previously noted by \citet{b33} can be
attributed to the orbital motion of the binary.

We have performed new multicolour polarimetric observations of HESS
J0632+057 and one nearby field star, revealing the precise wavelength
dependence of the source's intrinsic polarisation and confirming the
association between the variability of the polarisation and the
orbital phase of the binary.

\begin{table*}
 \centering
  \caption{Multicolour polarimetry for HESS J0632+057 from this work (Set "I" and Set "II") and polarimetric data of "UBV" monitoring early published by \citet{b5}.
  $\sigma_{1p}$ and $\sigma_{1\Theta}$ are the errors derived from photon counting statistics}
  \begin{tabular}{llccccccr}
  \hline
Source/JD &JD & Filter &Integation time, (s)  &  p$\pm\sigma$, (\%)&$\sigma_{1p}$ &$\Theta_{obs}\pm\sigma$    &$\sigma_{1\Theta}$ & p$_{circ}\pm\sigma$,  (\%)\\
\hline
 JD 2457106.0+&.28903603 & "U" &1593 & 3.43$\pm0.15$ &     $\pm0.08$ &    170.5$\pm3.0$ &     $\pm2.4$ &  0.051$\pm0.059$\\
(Set "I") &.26439721&  "B"&2500&3.92$\pm0.15$&  $\pm0.02$ & 171.7$\pm3.0$&$\pm0.7$&-0.003$\pm0.017$\\
 &.25615727& "V"& 700& 4.21$\pm0.15$&   $\pm0.04$&  172.3$\pm3.0$&
 $\pm1.1$&-0.054$\pm0.027$\\
 &.25017405& "R"& 292&3.83$\pm0.15$&   $\pm0.03$&  171.3$\pm3.0$&   $\pm0.9$&
 0.039$\pm0.021$\\
&  .24407511& "I"&309 & 3.78$\pm0.15$ &  $\pm0.03$&
172.0$\pm3.0$&
$\pm0.9$&  0.046$\pm0.022$  \\
 &  &&&&&&&\\
JD 2457369.0+&      .44839849&    "U"& 989  & 3.22$\pm0.15$&
$\pm0.09$&
166.4$\pm3.0$&     $\pm2.5$&   -0.173$\pm0.062$\\
(Set "II")& .45996389 & "B"& 400&3.78$\pm0.15$&  $\pm0.05$&
167.2$\pm3.0$&
$\pm1.4$&-0.014$\pm0.034$\\
&  .44659546 & "V"& 425 &3.87$\pm0.15$&   $\pm0.04$ &
167.5$\pm3.0$&
$\pm1.3$&-0.059$\pm0.031$\\
& .46968639& "R"&300 &3.74$\pm0.15$&   $\pm0.03$&  167.9$\pm3.0$&
$\pm0.8$& -0.013$\pm0.019$\\
&  .47327445& "I"& 400 &3.53$\pm0.15$&  $\pm0.03$&  170.1$\pm3.0$&
$\pm0.7$&  0.012$\pm0.018$  \\
\hline
\citet{b5}& &&&&&&&\\
 JD 2449745.0+&.36050&    "U" &240 &3.65$\pm$0.18&&161.9$\pm$5.1&$\pm$1.4&-0.080$\pm$ 0.130\\
&.36260&    "B" &180 &3.66$\pm$0.20&&165.6$\pm$5.8&$\pm$1.7&0.130$\pm$ 0.140\\
&.36400&    "V" &120 &3.82$\pm$0.11&&163.6$\pm$3.0&$\pm$0.8&-0.100$\pm$ 0.080\\
&  &&&&&&\\
JD 2449751.0+&.38290&    "V" &180& 3.87$\pm$0.09&&164.5$\pm$2.6&$\pm$0.7&0.080$\pm$ 0.060\\
&.38850&    "U" &240 &3.78$\pm$0.18&&163.2$\pm$5.1&$\pm$1.3&0.060$\pm$ 0.130\\
&.39060&    "B" &180 &3.80$\pm$0.20&&165.5$\pm$5.8&$\pm$1.6&-0.040$\pm$ 0.130\\
&.39270&    "V" &180 &3.80$\pm$0.09&&165.3$\pm$2.5&$\pm$0.6&0.010$\pm$ 0.060\\
&  &&&&&&\\
JD 2449753.0+&.30580&    "U"&240 & 3.75$\pm$0.20&&163.0$\pm$5.7&$\pm$1.5&-0.050$\pm$ 0.140\\
&.30790&    "B" &180 &3.63$\pm$0.23&&165.6$\pm$6.5&$\pm$1.8&-0.310$\pm$ 0.160\\
&.30990&    "V" &180 &4.15$\pm$0.10&&164.6$\pm$2.9&$\pm$0.5&0.060$\pm$ 0.070\\
&.31830&    "U" &240 &3.61$\pm$0.19&&163.9$\pm$5.5&$\pm$1.5&0.090$\pm$ 0.130\\
&.32040&    "B" &180 &3.87$\pm$0.21&&165.6$\pm$6.1&$\pm$1.6&-0.130$\pm$ 0.150\\
&.32240&    "V" &180 &3.58$\pm$0.09&&164.4$\pm$2.6&$\pm$0.7&-0.060$\pm$ 0.060\\
&.33010&    "U" &240 &3.66$\pm$0.19&&161.7$\pm$5.4&$\pm$1.5&-0.410$\pm$ 0.130\\
&.33220&    "B" &180 &4.00$\pm$0.22&&165.3$\pm$6.1&$\pm$1.5&-0.010$\pm$ 0.150\\
&.33430&    "V" &180 &3.90$\pm$0.09&&164.3$\pm$2.6&$\pm$0.7&-0.010$\pm$ 0.070\\
&.34330&    "U" &240 &3.66$\pm$0.19&&164.1$\pm$5.3&$\pm$1.5&0.130$\pm$ 0.130\\
&.34540&    "B" &180 &3.86$\pm$0.21&&161.9$\pm$6.0&$\pm$1.6&0.370$\pm$ 0.150\\
&.34740&    "V" &180 &3.74$\pm$0.09&&163.9$\pm$2.6&$\pm$0.7&-0.010$\pm$ 0.060\\
&.35780&    "U" &240 &3.82$\pm$0.19&&165.0$\pm$5.4&$\pm$1.4&-0.020$\pm$ 0.130\\
&.35990&    "B" &180 &3.65$\pm$0.21&&165.5$\pm$6.0&$\pm$1.7&0.120$\pm$ 0.150\\
&.36200&    "V" &180 &3.81$\pm$0.09&&164.4$\pm$2.6&$\pm$0.7&0.070$\pm$ 0.060\\
&.36970&    "U" &240 &3.52$\pm$0.19&&161.2$\pm$5.4&$\pm$1.5&0.070$\pm$ 0.130\\
&.37180&    "B" &180 &3.89$\pm$0.22&&169.2$\pm$6.1&$\pm$1.6&-0.140$\pm$ 0.150\\
&.37390&    "V" &180 &3.60$\pm$0.09&&165.3$\pm$2.6&$\pm$0.7&0.000$\pm$ 0.070\\
&.38150&    "U" &240 &3.53$\pm$0.19&&165.8$\pm$5.4&$\pm$1.5&-0.110$\pm$ 0.130\\
&.38360&    "B" &180 &3.63$\pm$0.21&&165.1$\pm$6.1&$\pm$1.7&0.180$\pm$ 0.150\\
&.38570&    "V" &180 &3.77$\pm$0.09&&165.1$\pm$2.6&$\pm$0.7&-0.100$\pm$ 0.070\\
&.39460&    "U" &240 &3.63$\pm$0.19&&165.7$\pm$5.4&$\pm$1.5&0.060$\pm$ 0.130\\
&.39670&    "B" &180 &3.61$\pm$0.22&&168.5$\pm$6.1&$\pm$1.7&0.160$\pm$ 0.150\\
&.39880&    "V" &180 &3.72$\pm$0.09&&163.9$\pm$2.6&$\pm$0.7&-0.110$\pm$ 0.070\\
&.40580&    "U" &240 &3.37$\pm$0.19&&162.7$\pm$5.5&$\pm$1.5&0.120$\pm$ 0.140\\
&.40780&    "B" &180 &3.95$\pm$0.22&&164.8$\pm$6.1&$\pm$1.6&-0.240$\pm$ 0.150\\
&.40990&    "V" &180 &3.78$\pm$0.09&&164.7$\pm$2.6&$\pm$0.7&-0.040$\pm$ 0.070\\
&.41760&    "U" &240 &3.39$\pm$0.19&&164.7$\pm$5.4&$\pm$1.6&-0.120$\pm$ 0.140\\
&.41970&    "B" &180 &3.70$\pm$0.22&&167.8$\pm$6.2&$\pm$1.7&-0.100$\pm$ 0.150\\
&.42170&    "V" &180 &3.77$\pm$0.09&&164.9$\pm$2.6&$\pm$0.7&-0.120$\pm$ 0.060\\
&.42940&    "U" &240 &3.13$\pm$0.19&&166.5$\pm$5.5&$\pm$1.6&0.110$\pm$ 0.140\\
&.43150&    "B" &180 &3.81$\pm$0.22&&164.5$\pm$6.2&$\pm$1.7&0.280$\pm$ 0.150\\
&.43360&    "V" &180 &3.56$\pm$0.09&&166.1$\pm$2.6&$\pm$0.7&-0.030$\pm$ 0.070\\
&  &&&&&&&\\
JD 2449755.0+&.28690&    "U"&240 & 3.54$\pm$0.21&&165.3$\pm$6.0&$\pm$1.7&0.100$\pm$ 0.150\\
&.28900&    "B" &180 &3.74$\pm$0.26&&166.3$\pm$7.3&$\pm$2.0&-0.270$\pm$ 0.180\\
&.29110&    "V" &180 &3.80$\pm$0.10&&164.1$\pm$2.6&$\pm$0.7&0.000$\pm$ 0.070\\
&.30700&    "U" &240 &3.41$\pm$0.19&&164.2$\pm$5.4&$\pm$1.6&-0.020$\pm$ 0.130\\
&.30910&    "B" &180 &3.72$\pm$0.22&&166.5$\pm$6.2&$\pm$1.7&0.160$\pm$ 0.150\\
&.31120&    "V" &180 &3.89$\pm$0.09&&165.0$\pm$2.6&$\pm$0.6&0.060$\pm$ 0.060\\
&.31810&    "U" &240 &3.66$\pm$0.19&&164.2$\pm$5.5&$\pm$1.5&0.380$\pm$ 0.140\\
&.32020&    "B" &180 &3.65$\pm$0.22&&164.6$\pm$6.3&$\pm$1.7&0.040$\pm$ 0.160\\
&.32230&    "V" &180 &3.82$\pm$0.09&&164.6$\pm$2.6&$\pm$0.7&0.040$\pm$ 0.070\\
\hline
\end{tabular}
\end{table*}

\begin{figure*}
\includegraphics[width=170mm, angle=0]{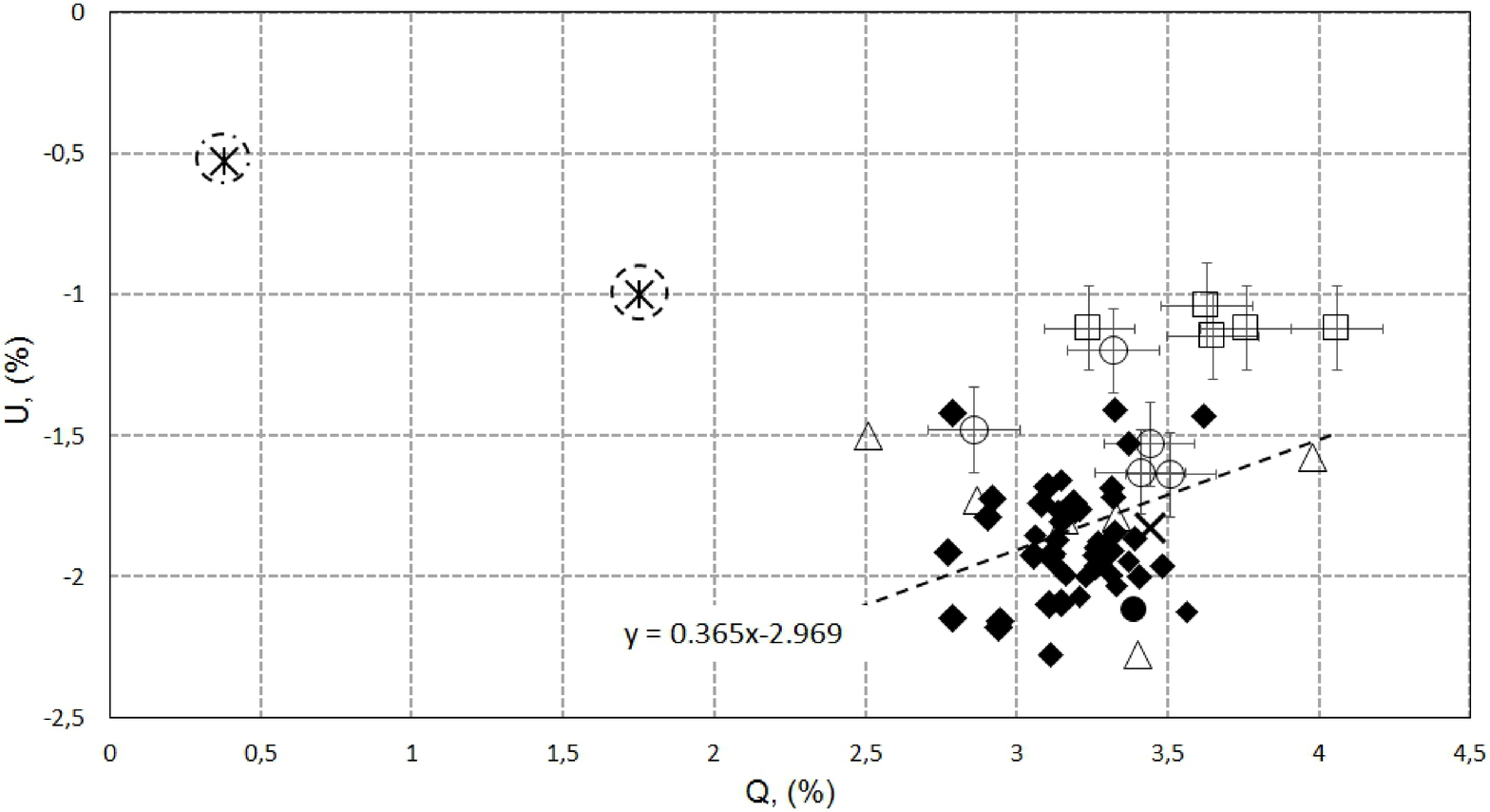}
 \vspace{-0.5cm}
 \caption{ qu-diagram of Stokes' parameters for
HESS J0632+057. Data from \citet{b5} - filled diamonds for $U\!BV$
bands. Data from \citet{b16}: open point-up triangles. Data from
\citet{b2}: cross. Data from \citet{b15}: filled circle. Data from
this work: open squares (Set I, JD2457106) and open circles (Set
II, JD2457369). Positions of interstellar polarisation (in V-band)
taken from \citet{b33} and from this work are indicated by dashed
and dashed-dotted circles respectively. Dashed line on the plot
represent the linear fit for the unweighted data as a whole.}
 \label{figure_2}
\end{figure*}

\section[]{Observations}
Multicolour optical polarimetry (linear and circular) of HESS
J0632+057 was carried out in two Sets (Set "I": 2015 March 24 (JD
2457106) and Set "II": 2015 December 12 (JD 2457369)) at the South
African Astronomical Observatory (SAAO). SAAO's high speed
polarimeter, HIPPO \citep{b40,b41}, was used to conduct all Stokes
mode photopolarimetry of HESS J0632+057 on the SAAO 1.9m
telescope. Observations were made with Johnson $U\!BVRI$ filters.
In addition, observations of a field star (GSC 00158-01500) were
performed in the same mode during JD 2457106. This A0V star
(m$_{V}$\,$\approx$\,12.1; "star in cluster") is located very
close to HESS J0632+057 (2 arcmin) and was used to estimate the
interstellar polarisation. The FK5 coordinates (ep=J2000 eq=2000)
for GSC 00158-01500 are 06 32 55.51 +05 49 29.9, while for HESS
J0632+057 the coordinates are 06 32 59.254 +05 48 01.18. The
resulting polarisation measurements for HESS J0632+057 are
presented in Table~2.

Assuming the orbital phase 0 is at T$_{0}$=MJD 2454857
\citep{b20}, the date of Set "I" observations correspond to an
orbital phase of 0.140 (if the period is 315 days), or phase 0.006
(if the period is 321 days) (see \citet{b20}). The Set "II"
observations fall at a phase of 0.975 (for 315 days), or 0.826
(for 321 days). In either case, our observations were performed
close to periastron passage - just after for Set "I" and just
before (the subsequent periastron passage) for Set "II".

\section[]{Discussion}
\subsection[]{Observed polarisation in HESS J0632+057}
As we noted in Sect.\,1, previously published polarimetric data
for HESS J0632+057 are very limited. The first observations made
by \citet{b15} at wavelength $\lambda=5400\AA$ (i.e. close to the
  photometric band V) observed polarisation to be p=4.0 per cent at
$\Theta=164^{o}$. Later, values of p=3.9 per cent at
$\Theta=166^o$ were published by \citet{b2} without any indication
of the
  wavelength used. We will assume further, that this data corresponded
  to the V-band.  The first multi-band measurements of polarisation
were presented by \citet{b16}, but with large errors in the red
($\sigma_{p}>$0.2 per cent at wavelengths $\sim$0.8-0.9$\mu$m). For
these observations, the position angle ($\Theta$) was also found to be
between $163^o$ to $166^o$. The most recent polarimetric data were
published by \citet{b5} and showed polarisation ranging from
p$\sim$3.5 per cent to p$\sim$3.9 per cent with $\Theta\sim162^{o}$ to
$166^{o}$ (see Table~2).

As shown by our Set "I" data, the object displayed statistically
higher polarisation than all previous measurements ($p_{V}=$4.21
per cent). In addition, the value of detected polarisation in the
red ($p_{I}\sim$3.8 per cent) and in the blue ($p_{U}\sim$3.4 per
cent) were around 0.5 per cent higher than noted by \citet{b16}.
At the same time, the position angle
($\Theta_{obs}\sim171^{o}-172^{o}$) is also different by
$\sim$6$^{o}$--10$^{o}$ from previously published values. The
polarised standard star, HD154445 \citep{b42,b43} was observed in
order to calculate the position angle offsets and efficiency
factors. Unpolarised standard stars were used to measure any
instrumental polarisation (0.4 per cent, 0.3 per cent, $<$0.1 per
cent, $<$0.1 per cent, $<$0.1 per cent in UBVRI correspondingly).
These values of instrumental polarisation were vectorially
subtracted from the observation values. Background sky
polarisation measurements were also taken at frequent intervals
during the observations. Thus, we can confidently rule out any
instrumental effects and conclude that the increase in
polarisation and shift in the observed position angle are real and
intrinsic to the source.

Our Set "II" data show statistically lower polarisation values (by
around 0.2-0.4 per cent) in all photometric bands compared to Set
"I" (see Fig.~1 and Table~2). The position angle also changed from
$\sim172^{o}$ (for Set "I") to $\sim167^{o}$.

The wavelength dependence of the observed polarisation for both of
our datasets was also different from that of \citet{b16} (see
Fig.~1). According to \citet{b16} the maximum polarisation
occurred at $\lambda$=0.62$\mu$m, while in our data the maximum is
in the "V"-band for both of our datasets.

The difference between our data and that previously published is
even more pronounced on a qu-diagram of Stokes' parameters
(Fig.~2). It is immediately obvious that our Set "I" data are
concentrated well outside the group of previously published
values. The Set "II" data are located in the space between Set "I"
and all other previously published values.

It is seen also in Fig.~2, that the data as a whole are spread out
around a straight line in the qu-plane although the correlation
coefficient is not large: r=0.33 for unweighted data (weighting
takes into account the observational errors changes the result
insignificantly). This may be linked to orbital phase, as a smooth
increase in polarisation the closer the compact object gets to
periastron. Note, however, that \citet{b33} found a very different
linear behaviour using only V-band data of \citet{b5}. Straight
lines in \citet{b33} and in Fig.~2 of the present work are
approximately perpendicular one to another in qu-plane. Better
time sampled polarimetric monitoring covering the orbital phase is
required to investigate this further. Finally, the analysis of
qu-diagram (Fig.~2) allow us to decline (at least to date) the
statement of \citet{b5} on the possibility that "the data trace an
ellipse in the qu-plane".

Generally, no significant circular polarisation was detected in either dataset. However, marginally significant ($\sim$2$\sigma$)
circular polarisation is seen in the "V" and "I" bands for Set "I" at the level $\sim$0.05 per cent, but with opposite signs.
Marginally significant ($\sim$2$\sigma$) circular polarisation was also seen in the "U" band for Set "II" at the level $\sim$0.17 per cent.
This result is in line with the data of \citet{b33}, who noted
that circular polarisation in HESS J0632+057 is very low, if it is present at all.

\begin{figure}
\includegraphics[width=80mm, angle=0]{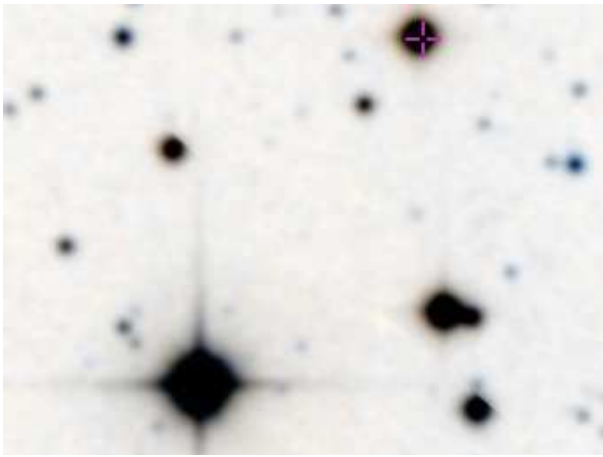}
\vspace{-0cm}
 \caption{Finding chart for the 2' region North-West of HESS J0632+057. The position of the observed field star GSC 00158-01500 is
indicated by the cross. The bright object in the lower left corner
is HESS J0632+057.} \label{figure_3}
\end{figure}

\subsection[]{Wavelength dependence of intrinsic polarisation in HESS J0632+057}
To date, we have two estimates of interstellar polarisation toward
the object. First is the estimate by \citet{b33} --
p$_{is}$\,$\approx$2 per cent and $\Theta_{is}\approx$165$^o$
(using the "field star method" for 28 field stars in 3$^{o}$
vicinity of HESS J0632+057) and second is the data obtained for
the nearby field star (GSC 00158-01500) in this work. This field
star is located just 2' from HESS J0632+057 (see Fig.~3) and was
not included in the previous interstellar polarisation estimate by
\citet{b33}. The nearest star from \citet{b46} catalogue used by
\citet{b33} for his interstellar polarisation estimate (HD 45910)
is located $\sim$40' away from HESS J0632+057. V-band measurements
obtained on the same date (JD 2457106.3413561) indicate the degree
of polarisation of p$_{V}=0.64\pm0.20$ per cent with
$\Theta\approx153^{o}\pm6^{o}$. Unfortunately, as this field star
is relatively faint (m$_{B,V,R}$\,$\approx$\,12.6;12.1;12.7
respectively), the detected degree of polarisation does not exceed
the 3$\sigma$ level in the "BRI" bands. Nevertheless, the degree
of polarisation detected in the "BRI" bands is less than 0.6 per
cent.

\begin{figure}
\includegraphics[width=90mm, angle=0]{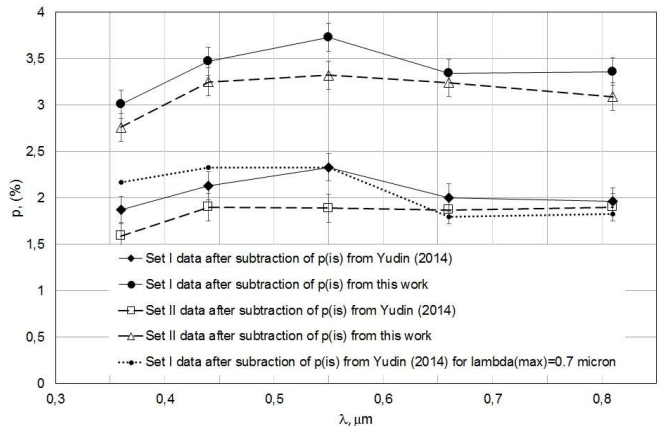}
\vspace{-1.0cm}
 \caption{Wavelength dependence of the polarisation intrinsic to HESS J0632+057.
 Connecting lines are not fits,
but are only the guide for eye. The solid lines represent the
wavelength dependence obtained from Set "I" after subtraction of
the interstellar polarisation taken from \citet{b33} (filled
diamonds) and this work (filled circles). The dashed lines
represent the wavelength dependence obtained from Set "II" after
subtraction of the interstellar polarisation taken from
\citet{b33} (open squares) and this work (open point-up
triangles). The dotted line represents the wavelength dependence
obtained from Set "I" after subtraction of the interstellar
polarisation taken from \citet{b33} when
$\lambda_{is}^{max}$=0.7$\mu$m was used instead of
$\lambda_{is}^{max}$=0.55$\mu$m.}   \label{figure_4}
\end{figure}
In order to obtain the wavelength dependence of the intrinsic
polarisation of the source, the interstellar component, p$_{is}$,
needs to be subtracted in all photometric bands. We can do this by
using both estimates of p$_{is}$ for our observing data of two
epochs reported above - Set I and Set II. For the data of
\citet{b33}, we assume that p$_{is}^{max}=$2 per cent,
$\Theta_{is}\approx$165$^o$ and $\lambda_{is}^{max}$=0.55$\mu$m.
The values of interstellar polarisation in other photometric bands
were calculated using the standard "Serkowski's Law" \citep{b38}.
Vectorial subtraction of these values of interstellar polarisation
from the observed ones allow us to re-construct the wavelength
dependence of the intrinsic polarisation (see Fig.~4). Note, that
only "V" band measurements for 28 field stars used by \citet{b33}
for the estimate of interstellar polarisation are available in
\citet{b46}. Therefore, some uncertainty in $\lambda_{is}^{max}$
value may occur. However, changing of $\lambda_{is}^{max}$ value
from 0.55$\mu$m to 0.7$\mu$m will change the resulting intrinsic
$p(\lambda)$ dependence insignificantly (see dotted line in Fig.~4
as an example).

Using this estimation of p$_{is}$, the average values of the
intrinsic polarisation for HESS J0632+057 are p$\approx$1.8 per
cent to 2.2 per cent and $\Theta\approx175^o\pm5^o$.

In Fig.~4 we also plot the wavelength dependence of the intrinsic
polarisation calculated by vectorial subtraction of the field star
observed during this work, using the above described procedure. We
assume that p$_{is}^{max}=$0.65 per cent,
$\Theta_{is}\approx$153$^o$ and $\lambda_{is}^{max}$=0.55$\mu$m.
In this case, the average values of the intrinsic polarisation for
HESS J0632+057 are p$\approx$3 per cent to 3.5 per cent and
$\Theta\approx173^o\pm4^o$. Note that in either case, the
intrinsic p($\lambda$) dependence is essentially flat, without
noticeably decreasing in the red. Similar p($\lambda$)
distributions are often observed in classical Be stars and are
attributed to scattering in flattened circumstellar gaseous discs
and/or jets around them (see also the polarimetric study of the
microquasar LS I +61$^o$303 by \citet{b44}). We propose that
Thomson scattering in a gaseous disc (see, for example, the
discussion of optical polarisation in the high-mass X-ray binary
LS 5039 by \citep{b39}) may be the dominant mechanism for
producing the polarisation behaviour in HESS J0632+057 (at least
close to periastron passage). Another possibility is discussed by
\citet{b34} in the context of synchrotron emission from jets
launched close to black holes or neutron stars in low-mass X-ray
binaries. In this case, however, strong polarimetric variability
can be expected in the near-IR.

\section{Conclusions}
We have shown that the observed polarisation in HESS J0632+057, close
to periastron passage, is statistically different from previously
published measurements. The detected polarisation degree, just after
periastron, is higher than previously observed with a shift of
6$^{o}$-10$^{o}$ in position angle. We propose that these changes
arise from the interaction of the compact object with a circumstellar
disc near periastron passage, and/or some redistribution of the
material within the circumstellar disc. According to \citet{b37}, the
compact object moves deeply through the circumstellar disc during
periastron passage, in the zones associated with $H_{\alpha}$,
$H_{\beta}$, $H_{\gamma}$ line formation, i.e. crossing dense regions
of the disc \citep{b35}. At this orbital phase, different events may
occur, such as: the interaction of the stellar wind with the wind or
jet of the compact object; a strong flow of matter from the Be star
can be captured by the compact object resulting in the formation of an
accretion disc or gaseous tail; the disruption of the circumstellar
disc by the compact object etc. (see \citet{b47}; \citet{b35};
\citet{b37}). Therefore, some perturbation of circumstellar material
at this orbital phase or the formation of an additional source of
polarisation cannot be surprising. On the other hand, the Be disc can
be misaligned with the orbital plane \citep{b45}. In this case, the
formation of a tilted accretion disc, formation of a jet which is
non-orthogonal to the Be disc, or formation of a tail from gas
captured by a compact object may result in a shift in the position
angle of the observed polarisation, as detected here.

If the polarimetric variations reported here (especially for Set
  "I") are linked to some form of perturbations in the circumstellar
  disc or the formation of an additional source of polarisation
  started just after the periastron passage, then we can expect
  further increase of polarisation at phases up to $\sim$0.3-0.5. In
  this context, the object exhibits a main outburst, at TeV and
  X-rays, at orbital phases $\sim$0.3 and a secondary lower outburst
  at orbital phases $\sim$0.7 \citep{b20}. Therefore, one might expect
  the geometry of the circumstellar envelope to also change during
  these phases.

Depolarisation across the H$\alpha$ line, which is often observed in
many early-type objects, can be used as a method for independently
deriving the interstellar polarisation \citep{b47,b48,b49} and
circumstellar envelope geometry. Therefore, spectropolarimetry around
H$\alpha$ would be very helpful.

Finally, higher-precision polarimetric measurements over the
orbital period are needed to study the HESS J0632+057 behavior in
the qu-plane. To date, our limited observations do not allow us to
make definitive conclusions about the geometry of the system, or
the mechanisms responsible for the detected polarimetric
variations. Most extensive polarimetric data were obtained by
\citep{b5} in "U,B,V" bands. Lack of polarimetric data in "R,I"
does not allow to analyze the object in a whole spectral region.
Near-IR polarimetric data would also be strongly desirable.

Therefore, additional multi-epoch and multicolour polarimetric
observations covering the entire orbit are strongly desirable to
verify potential orbital phase variations in the polarisation
degree and to map the geometry of the HESS J0632+057 system.

\section*{Acknowledgments}
We thank anonymous referee for very useful comments that improved
the presentation of the paper. This research has made use of the
SIMBAD database, operated at CDS, Strasbourg, France. LJT would
like to thank the Claude Leon Foundation and the University of
Cape Town Research Committee for their support. This material is
based upon work supported financially by the National Research
Foundation. Any opinions, findings and conclusions or
recommendations expressed in this material are those of the
author(s) and therefore the NRF does not accept any liability in
regard thereto.

\label{lastpage}

\end{document}